\documentclass[journal]{IEEEtran}

\usepackage{color}
\usepackage{cite}
\usepackage{amsthm}
\usepackage[cmex10]{amsmath}
\usepackage{algorithmic}
\usepackage{array}
\usepackage{mdwmath}
\usepackage{mdwtab}
\usepackage{amsfonts}
\usepackage{amssymb}
\usepackage{eqparbox}
\usepackage{multirow}
\usepackage{graphicx}
\usepackage[labelsep=period]{caption}
\usepackage{caption}
\usepackage{subcaption}
\usepackage{graphicx}
\usepackage{amsmath}

\usepackage{algorithm}
\usepackage{algorithmic}

\newcommand{\argmin}{\operatornamewithlimits{argmin}}
\newcommand{\argmax}{\operatornamewithlimits{argmax}}
\begin{document}
%--------------------------------------------------------------------------------------------------------%
\title{High Rate/Low Complexity Space-Time Block Codes for $2 \times 2$ Reconfigurable MIMO Systems}
%--------------------------------------------------------------------------------------------------------%
\author{Vida~Vakilian, Hani Mehrpouyan,~\IEEEmembership{Member,~IEEE,}
        Yingbo Hua, and~Hamid Jafarkhani,~\IEEEmembership{Fellow,~IEEE}%
\thanks{V. Vakilian is with the Dept. of Elect. and Comp. Engineering and Comp. Science, California State University, Bakersfield, CA 93311 USA (e-mail: vida.vakilian.ca@ieee.org).}
\thanks{H. Mehrpouyan is with the Dept. of Elect. and Comp. Engineering at the Boise State University, Boise, ID 83725 USA. (e-mail: hani.mehr@ieee.org).}
\thanks{Y. Hua is with the Dept. of Elect. Engineering at the University of California, Riverside, CA, 92521 USA. (e-mail: yhua@ee.ucr.edu).}
\thanks{H. Jafarkhani is with the Center for Pervasive Communications and Computing,
University of California, Irvine, CA 92617 USA (e-mail: hamidj@uci.edu).}}
%--------------------------------------------------------------------------------------------------------%
\maketitle
\begin{abstract}
In this paper, we propose a full-rate full-diversity space-time block code (STBC) for  $2 \times 2$ reconfigurable multiple-input multiple-output (MIMO) systems that require a low complexity maximum likelihood (ML) detector. We consider a transmitter equipped with a linear antenna array where each antenna element can be independently configured to create a directive radiation pattern toward a
selected direction\footnote{Composite right-left handed (CRLH) leaky-wave antenna (LWA) is an example of reconfigurable antennas with such characteristics \cite{lim2004electronically,caloz2008crlh}.}. This property of transmit antennas allow us to increase the data rate of the system, while reducing the computational complexity of the receiver. The proposed STBC achieves a coding rate of \textit{two} in a $2 \times 2$ MIMO system and can be decoded via an ML detector with a complexity of order $M$, where $M$ is the cardinality of the transmitted symbol constellation. Our simulations demonstrate the efficiency of the proposed code compared to existing STBCs  in the literature.
\end{abstract}
\begin{keywords}
Multiple-input multiple-output (MIMO), reconfigurable antennas, space-time coding.
\end{keywords}
%--------------------------------------------------------------------------------------------------------%
\section{Introduction}
\subsection{Motivation and Literature Survey}
\IEEEPARstart{S}{pace}-time block coding (STBC) technique is one of the most effective diversity methods used to combat the effect of channel fading in wireless communication \cite{alamouti1998simple,tarokh1999space,jafarkhani2001quasi,xian2005rate}. There are numerous studies on designing high-rate STBC for multiple-input multiple-output (MIMO) systems \cite{belfiore2005golden,IEEEStd2006Air,sezginer2007full,paredes20072a,rabiei2009new}. These codes are mostly designed based on the assumption that the antenna arrays at the transmitter and the receiver are omni-directional, i.e., there is no control mechanism over the signal propagation from each antenna element. Deploying reconfigurable antennas in MIMO arrays can add intelligence to these codes and provide additional degrees of freedom in the system that can be exploited to design STBCs with improved performance \cite{cetiner2006mimo,piazza2008design,frigon2008dynamic}. The design of such codes will be also useful for emerging wireless communication technologies such as millimeter wave (mmWave) systems, in which the use  of antennas with controllable radiation patterns is a necessity to overcome the severe path-loss and fading in high frequencies \cite{sayeed2013beamspace,vakilian2015STBC}.

Recently, several block-coding techniques have been designed to improve the performance of reconfigurable MIMO systems \cite{grau2008reconfigurable,fazel2009space,vakilian2013space}.
In~\cite{grau2008reconfigurable}, the authors proposed a coding scheme that can increase the diversity order of conventional MIMO systems by the number of the reconfigurable states at the receiver antenna. \cite{fazel2009space} extends the technique in~\cite{grau2008reconfigurable} to MIMO systems with reconfigurable antenna elements at both the transmitter and receiver sides, where a state-switching transmission scheme is used to further utilize the available diversity in the system over flat fading wireless channels. Later in \cite{vakilian2013space}, a coding scheme was proposed for reconfigurable MIMO systems over frequency-selective fading channels. However, using the aforementioned block codes in the system is only able to transmit one symbol per channel use, i.e., they do not provide any multiplexing gain. Moreover, the detection complexity of the codes is high and increases with the number reconfigurable states of the antenna array.

\subsection{Contributions}
In this paper, we propose a high-rate space-time block code for a $2 \times 2$ MIMO systems that are equipped with reconfigurable antenna elements. The proposed code uses the properties of the reconfigurable antennas to enhance the coding rate, while reducing the complexity of the maximum likelihood (ML) detector at the receiver. At each time slot, the system transmits multiple symbols over different beams each intended for a particular direction and receive antenna. We show that for a $2\times2$ MIMO system, a coding rate of \textit{two} can be achieved. Furthermore, due to the structure of the proposed code the system can use a conditional ML detection scheme to further reduce the complexity of the data detection process to ${\cal O}(M)$, where $M$ is the cardinality of the signal constellation, and ${\cal O}(\cdot)$ denotes the big omicron notation.

For comparison purposes, we study the performance of the recently developed rate-2 STBCs, including the \textit{Matrix C}~\cite{IEEEStd2006Air}, and \textit{maximum transmit diversity (MTD)}~\cite{rabiei2009new} codes. The Matrix C code is a threaded algebraic space-time code \cite{damen2003linear}, which is known as one of the well-performing STBCs for $2 \times 2$ MIMO systems. In \cite{rabiei2009new}, the authors proposed a high-rate STBC code, referred to as the MTD code, that is designed based on the linear combination of two Alamouti codes. The MTD code has an ML detection complexity of ${\cal O}(M^2)$. However, by taking advantage of the characteristics of reconfigurable antennas, our proposed code achieves an ML decoding complexity of ${\cal O}(M)$.

The rest of the paper is organized as follows. In Section~\ref{sec:proposedSTBC}, we introduce the proposed high-rate STBC for the $2 \times 2$ MIMO systems and describe the signal model. We describe the design criteria of the code and their relations with the parameters of the reconfigurable antennas in Section~\ref{sec:designcriteria}. We present a low complexity ML decoder for the proposed STBC in Section~\ref{sec:Decoding}. Simulation results are presented in Section~\ref{sec:results}, and concluding remarks are appeared in Section~\ref{sec:conc}.

\textit{Notation:} Throughout this paper, we use capital boldface letters, $\mathbf{X}$, for matrices and lowercase boldface letters, $\mathbf{x}$, for vectors. $(\cdot)^T$ denotes transpose operator. ${\bf A}\circ {\bf B}$ denotes the Hadamard product of the matrices ${\bf A}$ and ${\bf B}$, $||{\bf A}||_F$ represents the Frobenius norm of the matrix ${\bf A}$, $\text{det}({\bf A})$ computes the determinant of the matrix ${\bf A}$, and vec({\bf A}) denotes the vectorization of a matrix {\bf A} by stacking its columns on top of one another. Moreover, $\text{diag}(a_1, a_2, \cdots, a_n)$ represents a diagonal $n \times n$ matrix, whose diagonal entries are $a_1, a_2, \cdots , a_n$. ${\bf I}_M$ is the identity matrix of size $M$, and $Q(\cdot)$ is the Gaussian Q-function. We also use $\mathbb{C}$ to denote the set of complex valued numbers and $\text{Re}\{\cdot\}$ to represent the real component of a complex variable.

%--------------------------------------------------------------------------------------------------------%
\section{Proposed Space-Time Block Code}
\label{sec:proposedSTBC}
We consider a single-user mmWave system with $N_t  = 2$ and $N_r = 2$ transmit and receive antennas, respectively, where the transmitter is equipped with directive reconfigurable antennas to overcome the signal power degradation due to high pathloss in mmWave systems. We construct every $2\times 2$  codeword matrix from four information symbols  $\{s_1,s_2,s_3,s_4\}$  that will be sent during $T = 2$ time slots from $N_t = 2$ reconfigurable antennas. A $2 \times 2$ block code consist of four symbols is transmitted by $N_t = 2$ transmit antennas during $T = 2$ time slots, i.e.,
\begin{align}
{\bf C} &= \sqrt{P} \left[
\begin{array}{cc}
   c_1(1)&  c_2(1)   \\
   c_1(2)&  c_2(2)
\end{array}
\right] \nonumber\\ &= \sqrt{P}\left[
\begin{array}{cc}
   s_1 \alpha_1 - s_2^* \beta_1 &  s_3 \alpha_2 - s_4^* \beta_2  \\
  -s_3^* \alpha_2 + s_4 \beta_2 &  s_1^* \alpha_1 - s_2 \beta_1
\end{array}
\right],
\label{eq:code2tx2}
\end{align}
where $P$ is the transmit power per antenna, $\alpha_i = \sin(\theta_i)$, $\beta_i = \cos(\theta_i)$, for $i \in \{1, 2\}$. These choices for $\alpha_i$ and $\beta_i$ ensure that there is no transmit energy increase, i.e. $\alpha_i^2 + \beta_i^2 = 1$. In the next section, we explain how to choose optimal angles $\theta_1$ and $\theta_2$ to maximize the diversity and coding gains.

In order to reduce the decoding complexity, we can incorporate channel state information (CSI) and antenna radiation pattern characteristics into the block code at the transmitter. To do so, we multiply (\ref{eq:code2tx2}) by ${\bf \Psi}(\phi)$ which represents a matrix that is a function of the wireless channel coefficients ${\bf H}$ and antenna radiation pattern characteristics ${\bf G}(\phi)$. Note that in the time-division-duplex (TDD) systems, the CSI of the uplink can be used as the CSI of the downlink due to channel reciprocity \cite{marzetta2006fast} and, therefore, no receiver feedback is required. The proposed STBC then can be expressed as
\begin{align}
{\bf \boldsymbol{\cal C}} =\sqrt{P}\left[
\begin{array}{cc}
   s_1 \alpha_1 - s_2^* \beta_1 &  s_3 \alpha_2 - s_4^* \beta_2  \\
  -s_3^* \alpha_2 + s_4 \beta_2 &  s_1^* \alpha_1 - s_2 \beta_1
\end{array}
\right] \frac{{\bf \Psi}^H(\phi)}{||{\bf \Psi}(\phi)||_F},
\label{eq:code2tx2_Hg}
\end{align}
where the entries of ${\bf \Psi}(\phi) = [{\boldsymbol{\psi}}_1(\phi),\, {\boldsymbol{\psi}}_2(\phi)]$ can be computed as the Hadamard product of the channel matrix, ${\bf H}$, and the antenna gain matrix, ${\bf G}(\phi)$, i.e.,
\begin{align}
{\bf \Psi}(\phi) = {\bf H} \circ {\bf G}(\phi),
\end{align}
where ${\bf H} \triangleq [{\bf h}_{1},\; {\bf h}_{2}]$ with ${\bf h}_i  \triangleq [h_{i,1},\; h_{i,2}]^T$, and ${\bf G}(\phi) \triangleq [{\bf g}_1(\phi),\; {\bf g}_2(\phi)]$ with ${\bf g}_j(\phi) \triangleq [g_{j}(\phi_{1}),\, g_{j}(\phi_{2})]^T$.
If $N_s$ information symbols in a codeword are transmitted over $T$ channel uses, the transmission symbol rate is defined as $r_s  = \frac{N_s}{T}$, and the bit rate per channel use is computed as $r_b  = r_s \log_2 M$, where $M$ is the cardinality of the signal constellation. Note that a STBC is said to be full-rate when the number of transmitted symbols per channel use (pcu) is equal to the number of transmit antennas, i.e., when $r_s = N_t$  \cite{rabiei2009new}. Thus, based on this definition, the proposed STBC in (\ref{eq:code2tx2}) is full-rate ($r_s = N_t = 2$).

The overall received signal vector at the $i$-th antenna, ${\bf y}_i =[y_i(1), \, y_i(2)]^T$, can be written as 
\begin{align}
{\bf y}_i = \frac{\sqrt{P}}{||{\bf \Psi}(\phi)||_F}&\left[
\begin{array}{cc}
   s_1 \alpha_1 - s_2^* \beta_1 &  s_3 \alpha_2 - s_4^* \beta_2  \\
  -s_3^* \alpha_2 + s_4 \beta_2 &  s_1^* \alpha_1 - s_2 \beta_1
\end{array}\right] \nonumber \\ & \hspace{75pt} \times
\left[
\begin{array}{c}
 {\boldsymbol{\psi}}_1^H(\phi) {\boldsymbol{\psi}}_i(\phi)\\
 {\boldsymbol{\psi}}_2^H(\phi) {\boldsymbol{\psi}}_i(\phi)
\end{array}\right]+ {\bf z}_i,
\label{eq:receivedsignalmatrix}
\end{align}
where ${\boldsymbol{\psi}}_i(\phi) \triangleq [h_{i,1} g_{1}(\phi_{i}), \, h_{i,2} g_{2}(\phi_{i})]^T$ denotes the vector formed by the Hadamard product of the channel matrix and the antenna gain matrix, and ${\bf z}_i \triangleq [z_i(1), \, z_i(2)]^T$ represents the noise vector.
%--------------------------------------------------------------------------------------------------------%
\setcounter{equation}{15}
\begin{figure*}[!t]
\begin{align}
\theta^o_1 = \text{arctan} \sqrt{\frac{(|d_2|^2 + |d_3|^2)^2 - 2\Big(\big(\text{Re}\{d_1 d_2\} + \text{Re}\{d_3 d_4\}\big)^2 -\frac{1}{2} (|d_1|^2 + |d_4|^2)(|d_2|^2 + |d_3|^2)\Big)}{(|d_1|^2 + |d_4|^2)^2 - 2\Big(\big(\text{Re}\{d_1 d_2\} + \text{Re}\{d_3 d_4\}\big)^2 -\frac{1}{2} (|d_1|^2 + |d_4|^2)(|d_2|^2 + |d_3|^2)\Big)}},
\label{eq:alpha_beta3}
\end{align}
\end{figure*}
%--------------------------------------------------------------------------------------------------------%

%--------------------------------------------------------------------------------------------------------%
\section{Design Criteria}
\label{sec:designcriteria}
In this section, we present the procedure of finding the optimal values for the coefficients $\alpha_i$, $\beta_i$, and $g_j(\phi_i)$ that leads to maximum achievable diversity and coding gains for the proposed code. Let us denote two distinct sets of symbols by $\{s_1, s_2, s_3, s_4\}$ and $\{u_1, u_2, u_3, u_4\}$ and construct two distinct STBC codewords ${\bf \boldsymbol{\cal C}}$ and ${\bf \boldsymbol{\cal U}}$ using equation (\ref{eq:code2tx2_Hg}).
The two criteria that we use to design our code are: {\it Rank Criterion} and {\it Determinant Criterion} \cite{jafarkhani2005space}.

\noindent{\it Rank Criterion or Diversity Criterion}: To guarantee full diversity, matrix $({\bf \boldsymbol{\cal C}}-{\bf \boldsymbol{\cal U}})^H({\bf \boldsymbol{\cal C}}-{\bf \boldsymbol{\cal U}})$ over all pairs of distinct codewords must be full rank. We can smplify the rank criterion formulation using the determinant operation and state it as the following condition:
\setcounter{equation}{4}
\begin{align}
\text{det}\big[({\bf \boldsymbol{\cal C}}-{\bf \boldsymbol{\cal U}})^H({\bf \boldsymbol{\cal C}}-{\bf \boldsymbol{\cal U}})\big] \neq 0, \quad \text{for} \; {\bf \boldsymbol{\cal C}} \neq {\bf \boldsymbol{\cal U}}.
\label{eq:rank}
\end{align}
It can be verified that
\begin{align}
\text{det}&\big[({\bf \boldsymbol{\cal C}}-{\bf \boldsymbol{\cal U}})^H({\bf \boldsymbol{\cal C}}-{\bf \boldsymbol{\cal U}})\big] = \big(|D|^2+|D'|^2\big)^2 \nonumber \\
& \times \big|h_{1,1}g_1(\phi_1)h_{2,2}g_2(\phi_2)-h_{1,2}g_2(\phi_1)h_{2,1}g_1(\phi_2)\big|^2,
\label{eq:rank2}
\end{align}
where
\begin{align}
D &= d_1 \alpha_1 - d_2^* \beta_1,\label{eq:rotation3}\\
D' &= d_3 \alpha_2 - d_4^* \beta_2,
\label{eq:rotation4}
\end{align}
and $d_i = s_i - u_i$, for $i \in \{1, 2, 3, 4\}$. In order to achieve full-diversity, it is necessary to ensure that the coefficients $\alpha_i$, $\beta_i$, and $g_j(\phi_i)$ are chosen such that (\ref{eq:rank}) is greater than zero.

\noindent {\it Determinant Criterion or Coding gain Criterion}: The minimum coding gain distance between two distinct STBC codewords ${\bf \boldsymbol{\cal C}}$ and ${\bf \boldsymbol{\cal U}}$ can be expressed as
\begin{align}
\sigma_{min}= \min_{{\bf \boldsymbol{\cal C}} \neq {\bf \boldsymbol{\cal U}}} \text{det}\big[({\bf \boldsymbol{\cal C}}-{\bf \boldsymbol{\cal U}})^H({\bf \boldsymbol{\cal C}}-{\bf \boldsymbol{\cal U}})\big].
\label{eq:CGD1}
\end{align}
We need to be design the code such that the minimum coding gain distance, $\sigma_{min}$, is maximized. Let $\sigma$ denotes the achievable coding gain. We now can use (\ref{eq:CGD1}) to obtain $\sigma$ via the following optimization problem:
\begin{align}
\sigma &=\max_{\substack{g_{j}(\phi_i), \alpha_i, \beta_i \\ i, j \in \{1,2\}}} \; \min_{{\bf \boldsymbol{\cal C}} \neq {\bf \boldsymbol{\cal U}}} \text{det}\big[({\bf \boldsymbol{\cal C}}-{\bf \boldsymbol{\cal U}})^H({\bf \boldsymbol{\cal C}}-{\bf \boldsymbol{\cal U}})\big],\\
&= \max_{\substack{g_{j}(\phi_i), \alpha_i, \beta_i \\ i, j \in \{1,2\}}} \Big\{\min_{{\bf \boldsymbol{\cal C}} \neq {\bf \boldsymbol{\cal U}}} \;  \big(|D|^2+|D'|^2\big)^2 \nonumber \\
& \qquad   \times \big|h_{1,1}g_1(\phi_1)h_{2,2}g_2(\phi_2)-h_{1,2}g_2(\phi_1)h_{2,1}g_1(\phi_2)\big|^2\Big\},
\label{eq:CGD2}
\end{align}
Since the expression $\big(|D|^2+|D'|^2\big)^2$ only depends on $\alpha_i$, $\beta_i$, and the difference of the information symbols in the two distinct codewords, and the expression $\big|h_{1,1}g_1(\phi_1)h_{2,2}g_2(\phi_2)-h_{1,2}g_2(\phi_1)h_{2,1}g_1(\phi_2)\big|^2$ only depends on the channel coefficients, $h_{i,j}$, and antenna parameters, $g_j(\phi_i)$, the maximization problem in (\ref{eq:CGD2}) can be decoupled and rewritten as
\begin{align}
\sigma &= \max_{\substack{\alpha_i, \beta_i \\ i, j \in \{1,2\}}} \Big\{\min_{{\bf \boldsymbol{\cal C}} \neq {\bf \boldsymbol{\cal U}}} \;  \big(|D|^2+|D'|^2\big)^2\Big\} \nonumber \\
&  \times  \max_{\substack{g_{j}(\phi_i) \\ i, j \in \{1,2\}}} \Big\{\big|h_{1,1}g_1(\phi_1)h_{2,2}g_2(\phi_2)\nonumber \\ &\hspace{60pt}-h_{1,2}g_2(\phi_1)h_{2,1}g_1(\phi_2)\big|^2\Big\},
\label{eq:CGD3}
\end{align}
In sequel, we will explain how to find the coefficients $\alpha_i$, $\beta_i$, and $g_j(\phi_i)$ that maximize the minimum coding gain distance.
%--------------------------------------------------------------------------------------------------------%
\subsection{Optimal $\alpha_i$ and $\beta_i$}
As mentioned in the previous section, we choose $\alpha_i = \sin(\theta_i)$, $\beta_i = \cos(\theta_i)$, for $i \in \{1, 2\}$, such that the transmit power constraint is satisfied. To find the optimal value for $\theta_i$, we form the following optimization problem:
\begin{align}
\{\theta^o_1, \theta^o_2\} &=\argmax_{\substack{\theta_1, \theta_2 \,\in \, [0, \pi/2]}} \; \min_{{\bf \boldsymbol{\cal C}} \neq {\bf \boldsymbol{\cal U}}} \Big|\text{det}\big[({\bf \boldsymbol{\cal C}}-{\bf \boldsymbol{\cal U}})\big]\Big|,\\
&= \argmax_{\substack{\theta_1, \theta_2 \,\in \, [0, \pi/2]}} \; \min_{{\bf \boldsymbol{\cal C}} \neq {\bf \boldsymbol{\cal U}}} \Big| |d_1|^2 \alpha_1^2 + |d_2|^2 \beta_1^2 + |d_3|^2 \alpha_2^2 \nonumber \\ &+ |d_4|^2 \beta_2^2 - 2\text{Re}\{d_1 d_2 \alpha_1 \beta_1\} - 2\text{Re}\{d_3 d_4 \alpha_2 \beta_2\}
\Big|,
\label{eq:alpha_beta}
\end{align}
Subject to the constraint that $\theta_1+\theta_2 = \pi/2$, (\ref{eq:alpha_beta}) will take the following form:
\begin{align}
\theta^o_1 &= \argmax_{\substack{\theta_1}} \; \min_{{\bf \boldsymbol{\cal C}} \neq {\bf \boldsymbol{\cal U}}} \Big| (|d_1|^2 + |d_4|^2) \sin^2(\theta_1) \nonumber \\ &\hspace{60pt}+ (|d_2|^2 + |d_3|^2) \cos^2(\theta_1) \nonumber \\&\hspace{60pt}- \big(\text{Re}\{d_1 d_2\} + \text{Re}\{d_3 d_4\}\big) \sin(2\theta_1)
\Big|.
\label{eq:alpha_beta2}
\end{align}
By differentiating (\ref{eq:alpha_beta2}) with respect to $\theta_1$ and setting the result to zero, we get the optimal value of $\theta_1$ as shown in top of the page.
%--------------------------------------------------------------------------------------------------------%
\subsection{Optimal Reconfigurable Antenna Parameters}
We now proceed to obtain the optimal antenna parameter using the second optimization term in (\ref{eq:CGD3}). We formulate a cost function in terms of $g_j(\phi_i)$, for $i \in \{1, 2\}$, as follows:
\small
\setcounter{equation}{16}
\begin{align}
F({\bf g})&=\max_{g_{j}(\phi_i)} \Big\{\big|h_{1,1}g_1(\phi_1)h_{2,2}g_2(\phi_2)-h_{1,2}g_2(\phi_1)h_{2,1}g_1(\phi_2)\big|^2\Big\},\nonumber \\
&= \max_{g_{j}(\phi_i)} \Big\{\big| k g_1(\phi_1)g_2(\phi_2)-g_2(\phi_1)g_1(\phi_2)\big|^2\Big\},
\label{eq:CGD4}
\end{align}
\normalsize
where ${\bf g} = [g_1(\phi_1), g_1(\phi_2), g_2(\phi_1), g_2(\phi_2)]$ and $k = h_{1,1}h_{2,2}/h_{1,2}h_{2,2}$. This is a nonlinear optimization problem which solving it in the current form may lead to several local maxima. To avoid such scenarios, we first assume that the first and second transmit antennas are configured such that
\begin{align}
g_{1}(\phi_1) = g_{2}(\phi_2),\\
g_{1}(\phi_2) = g_{2}(\phi_1).
\end{align}
Second, to model the antenna radiation pattern, we consider a rectangular function as a proper abstraction to capture the direction steerability and beamwidth characteristics of the antenna \cite{li2009capacity}.  Using this model, we have
\begin{align}
g_{1}(\phi_2) B_{3dB} + g_{1}(\phi_1) B_{3dB} = 2\pi,
\label{ant_gain}
\end{align}
where $B_{3dB}$ is the 3-dB antenna beamwidth. From (\ref{ant_gain}), we obtain
\begin{align}
g_{1}(\phi_2) = \frac{2\pi - g_{1}(\phi_1) B_{3dB}}{B_{3dB}}.
\label{ant_gain2}
\end{align}
Substituting (\ref{ant_gain2}) into (\ref{eq:CGD4}), we arrive at:
\small
\begin{align}
F \big(g_1(\phi_1)\big)= \max_{g_{1}(\phi_1) \in (0, g_{up}]} \Big\{\big| k g_1(\phi_1)^2-( \frac{2\pi - g_{1}(\phi_1) B_{3dB}}{B_{3dB}})\big|^2\Big\},
\label{ant_gain3}
\end{align}
\normalsize
where $g_{up} = \frac{\pi}{B_{3dB}}$ is the upper bound value of the antenna gain. The maximization over $g_{1}(\phi_1)$ can be performed by differentiating (\ref{ant_gain3}) with respect to $g_{1}(\phi_1)$ and setting the result to zero. This leads to the following solutions:
\begin{align}
g_{1}^{s_1}(\phi_1) &= \frac{-2\pi (1-\sqrt{k})}{B_{3dB}(k-1)},\\
g_{1}^{s_2}(\phi_1) &= \frac{-2\pi (1+\sqrt{k})}{B_{3dB}(k-1)}, \\
g_{1}^{s_3}(\phi_1) &= \frac{-2\pi}{B_{3dB}(k-1)}.
\label{ant_gain4}
\end{align}
The above solutions shows the local optimum points of the cost function $F \big(g_1(\phi_1)\big)$. To obtain a pattern of the cost function, we compute the second derivative of $F \big(g_1(\phi_1)\big)$ with respect to $g_1(\phi_1)$ at these local optimum points which can be given by
\begin{align}
\frac{\partial^2 F \big(g_1(\phi_1)\big)}{\partial g_1(\phi_1)^2}\Bigg|_{g_{1}^{s_1}(\phi_1)} = \frac{16\pi^2}{B_{3dB}}k,\label{ant_gain5}\\
\frac{\partial^2 F \big(g_1(\phi_1)\big)}{\partial g_1(\phi_1)^2}\Bigg|_{g_{1}^{s_2}(\phi_1)} = \frac{16\pi^2}{B_{3dB}}k,\label{ant_gain6}\\
\frac{\partial^2 F \big(g_1(\phi_1)\big)}{\partial g_1(\phi_1)^2}\Bigg|_{g_{1}^{s_3}(\phi_1)} = \frac{-8\pi^2}{B_{3dB}}k.
\label{ant_gain7}
\end{align}
%
%--------------------------------------------------------------------------------------------------------%
\begin{figure}
\centering
\includegraphics [width=\columnwidth]{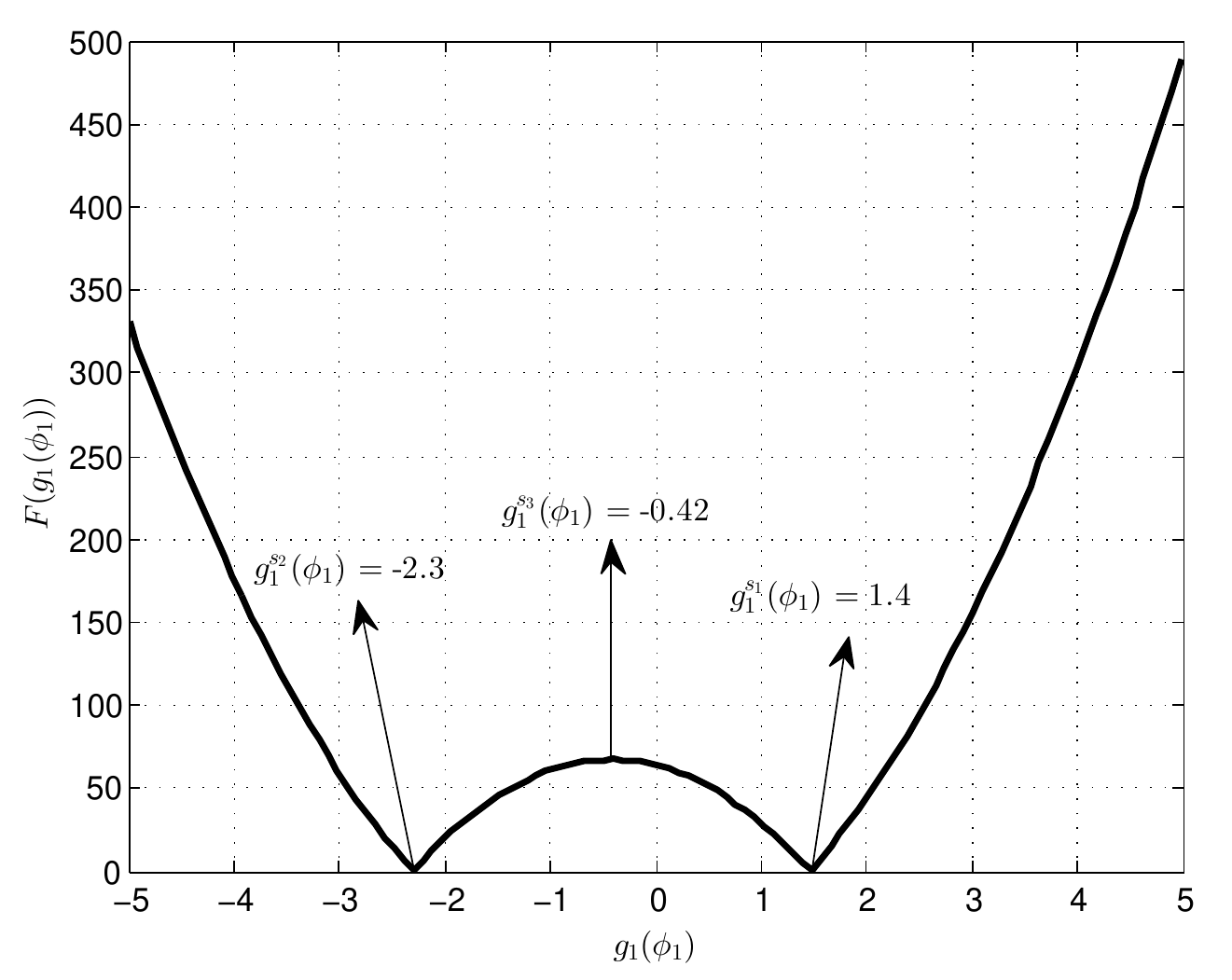}
\caption{\small Cost function $F \big(g_1(\phi_1)\big)$ versus different values of $g_1(\phi_1)$ for real positive $k$ .}
\label{fig:costfunc_2D_posk}
\end{figure}
%--------------------------------------------------------------------------------------------------------%
\begin{figure}
\centering
\includegraphics[width=\columnwidth]{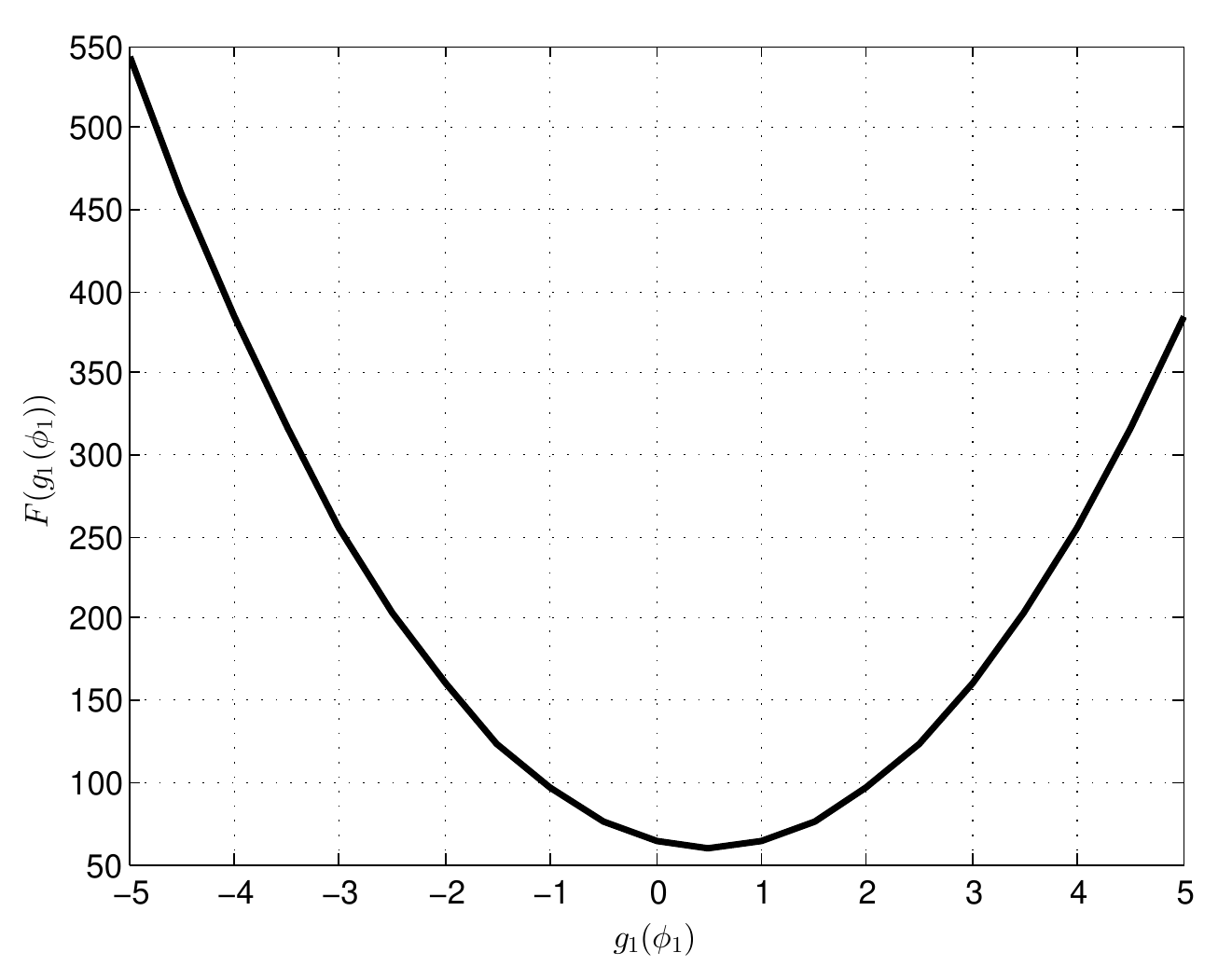}
\caption{\small Cost function $F \big(g_1(\phi_1)\big)$ versus different values of $g_1(\phi_1)$ for real negative $k$.}
\label{fig:costfunc_2D_negk}
\end{figure}
%--------------------------------------------------------------------------------------------------------%
%\begin{figure}
%\centering
%\includegraphics [width=\columnwidth]{./Figures/CostFunctionvsAntGain_k2+3i,B45_3D}
%\caption{\small Cost function $F \big(g_1(\phi_1)\big)$ versus different values of $g_1(\phi_1)$ for complex $k$.}
%\label{fig:costfunc_3D_complexk}
%\end{figure}
%--------------------------------------------------------------------------------------------------------%
\noindent From (\ref{ant_gain5})-(\ref{ant_gain7}), we observe that the local maxima and minima points of the cost function depends on $k$. Let us evaluate the cost function for two different sets of $k$ values. For the first case, we assume that $k$ takes positive real values, i.e., $k \in [0, \infty)$, and for the second case $k$ is assumed to be negative real. For the former, $g_{1}^{s_1}(\phi_1)$ and $g_{1}^{s_2}(\phi_1)$ are the global minimum points and $g_{1}^{s_3}(\phi_1)$ is  the local maximum. It can be verified that the cost function, $F (g_1(\phi_1))$, approaches positive infinity when $g_{1}(\phi_1)$ approaches to negative or positive infinity. Fig. \ref{fig:costfunc_2D_posk} shows the the cost-function curve behavior for $k = 20$ and $B_{3dB} = \pi/4$. For the latter, when $k$ is negative real,  the cost function will have a global minimum equal to $g_{1}^{s_3}(\phi_1)$. Fig. \ref{fig:costfunc_2D_negk} shows the pattern for negative real value of $k$, i.e., $k = -15$ and $B_{3dB} = \pi/4$, which is parabola shape as our calculation showed.
%--------------------------------------------------------------------------------------------------------%
\section{ML Decoding}
\label{sec:Decoding}
In this section, we formulate the ML decoding problem for the code proposed in reconfigurable MIMO systems. The decoder at the first and second branches respectively receive signals ${\bf y}_1$ and ${\bf y}_2$ during $T = 2$ time slots as computed in (\ref{eq:receivedsignalmatrix}). Assuming perfect channel state information at the receiver, the symbols $\{s_1, s_2, s_3, s_4\}$ can be jointly detected using an ML decoder as follows:
\begin{equation}
(\hat{s}_1, \hat{s}_2, \hat{s}_3, \hat{s}_4)=\argmin_{s_1, s_2, s_3, s_4}||{\bf y}-\widetilde{\bf \boldsymbol{\cal C}}{\bf h}_g||^2
\label{eq:joint-ML_total}
\end{equation}
\normalsize
where ${\bf y} = [{\bf y}_1^T, \, {\bf y}_2^T]^T$ is the overall received signal vector. The maximum likelihood decoder performs an exhaustive search over all possible values of the transmitted symbols and decides in favor of the quadruplet ($s_1, s_2, s_3, s_4$) that minimizes the Euclidean distance metric of (\ref{eq:joint-ML_total}). The computational complexity of the receiver in this case is ${\cal O}(M^4)$.  However, considering the structure of the proposed STBC code, the receiver without using CSI decouples symbol pairs $(s_1,s_2)$ and $(s_3,s_4)$ as
\small
\begin{align}
r_1 &= \frac{y_1(1) + y_2^*(2)}{\sqrt{2}}= \frac{\sqrt{P}}{\sqrt{2}||{\bf H}_g(\phi)||_F}\big[( s_1 \alpha_1 - s_2^* \beta_1){\bf h}_{g_1}^H{\bf h}_{g_1}\nonumber \\&+(s_3 \alpha_2 - s_4^* \beta_2){\bf h}_{g_2}^H{\bf h}_{g_1}-(s_3 \alpha_2 - s_4^* \beta_2){\bf h}_{g_2}^H{\bf h}_{g_1}\nonumber \\
&+(s_1 \alpha_1 - s_2^* \beta_1){\bf h}_{g_2}^H{\bf h}_{g_2}\big]+\frac{z_1(1)+z_2^*(2)}{\sqrt{2}},\nonumber \\
&= \sqrt{\frac{P}{2}}||{\bf H}_g(\phi)||_F(s_1 \alpha_1 - s_2^* \beta_1)+\frac{z_1(1)+z_2^*(2)}{\sqrt{2}}
\label{eq:r1}
\end{align}
\normalsize
and
\small
\begin{align}
r_2 &= \frac{y_2(1) - y_1^*(2)}{\sqrt{2}}= \frac{\sqrt{P}}{\sqrt{2}||{\bf H}_g(\phi)||_F}\big[(s_1 \alpha_1 - s_2^* \beta_1){\bf h}_{g_1}^H{\bf h}_{g_2}\nonumber \\&+(s_3 \alpha_2 - s_4^* \beta_2){\bf h}_{g_2}^H{\bf h}_{g_2}
+(s_3 \alpha_2 - s_4^* \beta_2){\bf h}_{g_1}^H{\bf h}_{g_1}\nonumber \\&-(s_1 \alpha_1 - s_2^* \beta_1){\bf h}_{g_1}^H{\bf h}_{g_2}\big]+\frac{z_2(1)-z_1^*(2)}{\sqrt{2}},\nonumber \\
&= \sqrt{\frac{P}{2}}||{\bf H}_g(\phi)||_F(s_3 \alpha_2 - s_4^* \beta_2)+\frac{z_2(1)-z_1^*(2)}{\sqrt{2}}
\label{eq:r2}
\end{align}
\normalsize
Hence, the ML metric in (\ref{eq:joint-ML_total}) can be broken into two independent sub-minimization problems as given below
\begin{equation}
(\hat{s}_1, \hat{s}_2)=\argmin_{s_1,s_2} |r_1-\sqrt{\frac{P}{2}}||{\bf H}_g(\phi)||_F(s_1 \alpha_1 - s_2^* \beta_1)|^2,
\label{eq:joint-ML-1}
\end{equation}
\begin{equation}
(\hat{s}_3, \hat{s}_4)=\argmin_{s_3, s_4} |r_2-\sqrt{\frac{P}{2}}||{\bf H}_g(\phi)||_F(s_3 \alpha_2 - s_4^* \beta_2)|^2.
\label{eq:joint-ML-2}
\end{equation}
Therefore, instead of minimizing the cost function in (\ref{eq:joint-ML_total}) over all possible values of ($s_1, s_2, s_3, s_4$), one can simultaneously minimize the cost functions in (\ref{eq:joint-ML-1}) and (\ref{eq:joint-ML-2}) over ($s_1, s_2$) and ($s_3, s_4$), respectively. Therefore, the ML decoding is realized by joint searches of two information symbols which results a complexity of ${\cal O}(M^2)$. As we will show in the following section, the ML decoding complexity of the proposed STBC can be further decreased to ${\cal O}(M)$.
%--------------------------------------------------------------------------------------------------------%
\subsection{Conditional ML Decoding}
\label{sec:MLConDecoding}
To reduce the decoding complexity of the proposed code, we use a conditional ML decoding technique \cite{sezginer2007full} as elaborated below.
Let us compute the following intermediate signals using the received signal, $r_1$, in~\eqref{eq:r1}, for a given value of the symbol $s_2$ 
\begin{align}
{\tilde r}_1 &=r_1-\sqrt{\frac{P}{2}}||{\bf H}_g(\phi)||_F(-s_2^* \beta_1) \nonumber\\
           &=  \sqrt{\frac{P}{2}}||{\bf H}_g(\phi)||_F(s_1 \alpha_1)+\frac{z_1(1)+z_2^*(2)}{\sqrt{2}}
\label{eq:r_1_1} 
%\\
%{\tilde r}_2 &= r_2+\sqrt{\frac{P}{2}}||{\bf H}_g(\phi)||_F(s_4^* \beta_2) \nonumber\\
%           &=  \sqrt{\frac{P}{2}}||{\bf H}_g(\phi)||_F(s_3 \alpha_2)++\frac{z_2(1)-z_1^*(2)}{\sqrt{2}}.
%\label{eq:r_2_1}
\end{align}
It can be seen from (\ref{eq:r_1_1}) that ${\tilde r}_1$ has only terms involving the symbol $s_1$ and, therefore, it can be used as the input signal to a threshold detector to get the ML estimate of the symbol $s_1$ conditional on $s_2$. As a result, instead of minimizing the cost function in (\ref{eq:joint-ML-1}) over all possible pairs $(s_1,s_2)$, we first obtain the estimate of $s_1$ using threshold detector, called $s_1^{ML}(s_2^{m})$, and then compute the cost function for ($s_1^{ML}(s_2^{m}),s_2^{m}$), for $m=1,2, \cdots, M$. The optimal solution can be obtained as 
\begin{equation}
\hat{s}_2=\argmin_{m} f\Big(s_1^{ML}(s_2^{m}),s_2^m\Big),
\label{eq:joint-ML-1_cond}
\end{equation}
where
\small
\begin{align}
f\Big(s_1^{ML}(s_2^{m}),&s_2^m\Big) = \;|r_1-\sqrt{\frac{P}{2}}||{\bf H}_g(\phi)||_F(s_1^{ML}(s_2^{m}) \alpha_1 - s_2^{m^*} \beta_1)|^2.\label{eq:costfunc1}
\end{align}
\normalsize
Similarly, $s_3$ and $s_4$ can be detected by solving the following 
\begin{equation}
\hat{s}_4=\argmin_{m} f\Big(s_3^{ML}(s_4^{m}),s_4^m\Big),
\label{eq:joint-ML-2_cond}
\end{equation}
where
\small
\begin{align}
f\Big(s_3^{ML}(s_4^{m}),&s_4^m\Big) = \;|r_2-\sqrt{\frac{P}{2}}||{\bf H}_g(\phi)||_F(s_3^{ML}(s_4^{m}) \alpha_2 - s_4^{m^*} \beta_2)|^2.\label{eq:costfunc2}
\end{align}
\normalsize
Using the above described conditional ML decoding, we reduce the ML detection complexity of the proposed code  from ${\cal O}(M^2)$ to ${\cal O}(M)$ (see Algorithm \ref{Alg1}).
\begin{algorithm}
\caption{Conditional ML Decoding $i\in \{1, 2\}$}
\label{Alg1}
{\bf Step 1:} Select $s_{2i}^m$ from the signal constellation set. \newline 
{\bf Step 2:} Compute ${\tilde r}_i$. \newline  
{\bf Step 3:} Supply ${\tilde r}_i$ into a phase threshold detector to get the estimate of $s_{2i-1}$ conditional on $s_{2i}^m$, called $s_{2i-1}^{ML}(s_{2i}^{m})$. \newline 
{\bf Step 4:} Compute the cost function in (\ref{eq:costfunc1}) for $s_{2i-1}^{ML}(s_{2i}^{m})$ and $s_{2i}^{m}$.\newline
{\bf Step 5:} Repeat Step 1 to Step 4 for all the remaining constellation points.\newline
{\bf Step 6:} The $s_{2i-1}^{ML}(s_{2i}^{m})$ and $s_{2i}^{m}$ corresponding to cost function with minimum value will be the estimate of $s_{2i-1}$ and $s_{2i}$.  \newline
\end{algorithm}
%--------------------------------------------------------------------------------------------------------%
\section{Simulation Results}
\label{sec:results}
In this section, we present the results of our numerical simulation to demonstrate the performance of the proposed coding scheme and compare it to the existing space-time coding methods in the literature. In particular, we compare the BER performance of the proposed code with the \textit{Dual Alamouti} \cite{li2011dual}, and \textit{MTD} \cite{rabiei2009new} schemes. Throughout the simulations, we assume a $2 \times 2$ MIMO structure. 
To investigate the effect of the phase noise, the received signal at the $n$-th antenna during the $k$-th time slot can be modeled as
\begin{eqnarray}
y_i(n) = \sum_{j=1}^{N_t} e^{\j\theta_i^{[r]}(n)} h_{i,j} g_{j}(\phi_{i}) e^{\j\theta_j^{[t]}(n)}{\cal C}_j(n)  + z_i(n),
\label{eq:Rx_Signal}
\end{eqnarray}
where $\j= \sqrt{-1}$ is the imaginary unit, $e^{\j\theta_j^{[t]}(n)}$ and $e^{\j\theta_j^{[r]}(n)}$ are used to respectively model the phase noise at the $j$-th transmit and the $i$-th receive antenna, ${\cal C}_j(n)$ is the entry of ${\bf \boldsymbol{\cal C}}$ as computed in (\ref{eq:code2tx2_Hg}) which is transmitted from the $j$-th transmit antenna during the $n$-th time slot, $h_{i,j}$ is the channel fading coefficient between the $j$-th transmit and the $i$-th receive antenna, $g_{j}(\phi_{i})$ is the antenna gain at the receive antenna $i$ from the transmit antenna $j$, and $z_i(n)$ is the additive Gaussian noise with zero-mean and variance $N_0$. The phase noise process can be modeled as a Brownian motion or Wiener process \cite{demir2000phase}
\begin{align}
e^{\j\theta_j^{[t]}(n)} &= e^{\j\theta_j^{[t]}(n-1)} + \Delta_j^{[t]}(n),\\
e^{\j\theta_i^{[r]}(n)} &= e^{\j\theta_i^{[t]}(n-1)} + \Delta_i^{[r]}(n)
\label{eq:Rician_ch}
\end{align}
where  $\Delta_j^{[t]}(n) \sim \mathcal{N}(0, \sigma^2_{\!\Delta_j^{[t]}})$ and $ \Delta_i^{[r]}(n) \sim \mathcal{N}(0, \sigma^2_{\!\Delta_j^{[t]}})$ are assumed to be white real Gaussian processes.  
We also consider Rician fading channel model with the following form
\begin{align}
h_{i,j}= \sqrt{\frac{K(f_c)}{s}\Big(\frac{d_0}{d}\Big)^\gamma} \Big(\sqrt{\frac{K_R}{K_R+1}} h_{i,j}^{L} + \sqrt{\frac{1}{K_R+1}} h_{i,j}^{N}\Big),
\label{eq:Rician_ch}
\end{align}
where 
\begin{itemize}
\item $K(f_c) \triangleq (\frac{\lambda}{4\pi d_0})^2$, $\lambda = \frac{c}{f_c}$ is the wavelength, $c$ is the speed of light, $f_c$ is the carrier frequency, $d_0$ is the refrence distance, $d$ is the distance between transmitter and receiver, and $\gamma$ is the path loss exponent.
\item $s$ is log-normally distributed random variable with mean $\mu_s$ and standard deviation $\sigma_s$ which models the shadowing effect.
\item $K_R$ is the Rician factor expressing the ratio of powers of the free-space signal and the scattered waves. 
\item Using this model, $h_{i,j}$ is decomposed into the sum of a random component, $h_{i,j}^N$ and the deterministic component $h_{i,j}^{L}$. The former accounts for the scattered signals with its entries being modeled as independent and identically distributed (i.i.d) complex Gaussian random variables with zero mean and unit variance. The latter, $h_{i,j}^{L}$, models the LoS signals.
\end{itemize}
%--------------------------------------------------------------------------------------------------------%
\begin{figure}
\centering
\includegraphics [width=\columnwidth]{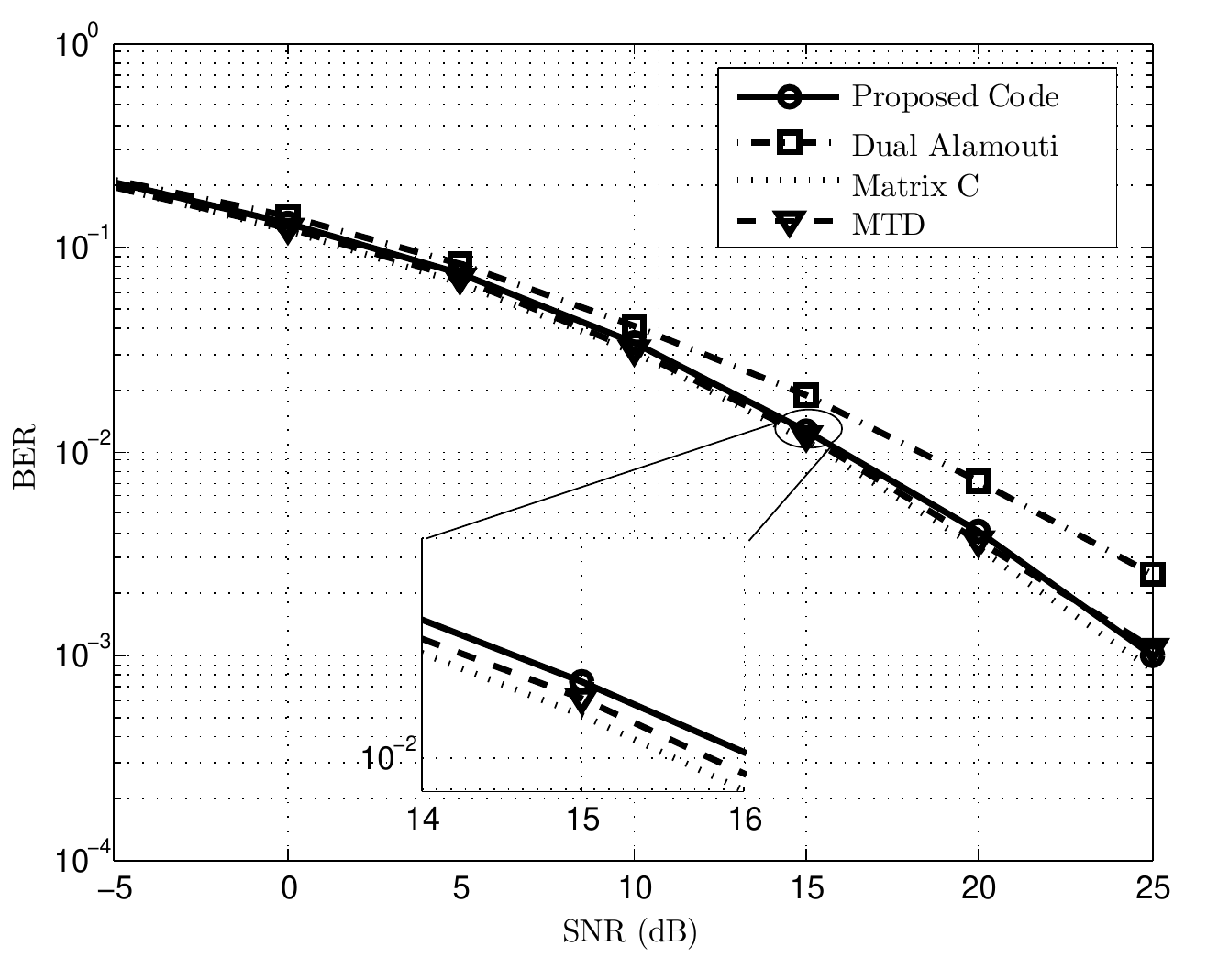}
\caption{\small BER performance of the proposed code with spectral efficiency of 4 bits pcu.}
\label{fig:ber-performance-4bcu}
\vspace{-10pt}
\end{figure}
\normalsize
%--------------------------------------------------------------------------------------------------------%
Fig. \ref{fig:ber-performance-4bcu} illustrates the BER performance of the proposed code in comparison with the performance of the Dual Alamouti, and MTD schemes over a Rician fading channel with K-factor equal to $5$ dB.
%--------------------------------------------------------------------------------------------------------%
\begin{table}
\caption{Simulation Parameters}
\begin{center}
\begin{tabular}{l||p{2.5cm}}
\hline\hline
 Parameters & Value \\ \hline\hline
 Carrier Frequency ($f_c$)                                                            & 60 GHz    \\
 Distance between transmitter and receiver ($d$)                         & 25 m         \\
 Path loss exponent ($\gamma$)                                                    & 4               \\
$\mu_s, \sigma_s$                                                                         & 0, 9            \\
$\sigma^2_{\!\Delta_j^{[t]}} = \sigma^2_{\!\Delta_j^{[r]}}$    & $3 \times 10^{-3}$ $\text{rad}^2$    \\  \hline \hline
\end{tabular}
\end{center}
\end{table}
%--------------------------------------------------------------------------------------------------------%

%--------------------------------------------------------------------------------------------------------%
\section{Conclusions}
\label{sec:conc}
We proposed a  full-rate full-diversity space-time code for wireless systems employing antennas with reconfigurable radiation patterns. Due to the structure of the proposed code and reconfigurable feature of the antenna elements, we reduce the  ML decoding complexity of the data detection to  ${\cal O}(M)$ which has significant impact on the energy consumption of the receiver especially for higher order modulation schemes. We provided simulation results that demonstrate the performance of the proposed code and made comparisons with that of the previous STBC coding schemes.  

\bibliographystyle{IEEEtran}
%{\footnotesize\bibliography{IEEEabrv,Reference}}
\bibliography{IEEEabrv,Reference}

\end{document}